\documentclass[manuscript,natbib=false]{acmart}
\AtBeginDocument{%
  }

\setcopyright{acmlicensed}
\copyrightyear{2026}
\acmYear{2026}
\acmDOI{XXXXXXX.XXXXXXX}
\acmConference[LAK'26]{the 16th International Conference on Learning Analytics and Knowledge}{April 27--May, 01, 2026}{Bergen, Norway}
\acmISBN{978-1-4503-XXXX-X/2018/06}

\usepackage{pdflscape}
\usepackage{booktabs, multirow, tabularx, array}

\newlength{\rqwidth}
\newlength{\rqindent}            
\begin{document}

\title{Investigating Self-regulated Learning Sequences within a Generative AI-based Intelligent Tutoring System}


\author{Jie Gao}
\affiliation{%
  \institution{McGill University}
  \city{Montreal}
  \state{Quebec}
  \country{Canada}}
\orcid{0000-0002-6933-950X}
\email{jie.gao3@mail.mcgill.ca}

\author{Shasha Li}
\affiliation{%
  \institution{McGill University}
  \city{Montreal}
  \state{Quebec}
  \country{Canada}}
\orcid{0009-0005-8343-3331}
\email{shasha.li@mail.mcgill.ca}

\author{Jianhua Zhang}
\affiliation{%
  \institution{The Chinese University of Hong Kong}
  \city{HongKong}
  \country{China}}
\orcid{0009-0002-8316-7545}
\email{1155203828@link.cuhk.edu.hk}

\author{Shan Li}
\affiliation{%
  \institution{Lehigh University}
  \city{Bethlehem}
  \state{Pennsylvania}
  \country{USA}}
\orcid{0000-0001-6001-1586}
\email{shla22@lehigh.edu}

\author{Tingting Wang}
\affiliation{%
  \institution{Renmin University of China}
  \city{Beijing}
  \country{China}}
\orcid{0000-0002-6315-4029}
\email{tingtingwang2024@ruc.edu.cn}

\renewcommand{\shortauthors}{Gao et al.}

\begin{abstract}
  There has been a growing trend in employing generative artificial intelligence (GenAI) techniques to support learning. Moreover, scholars have reached a consensus on the critical role of self-regulated learning (SRL) in ensuring learning effectiveness within GenAI-assisted learning environments, making it essential to capture students' dynamic SRL patterns. In this study, we extracted students’ interaction patterns with GenAI from trace data as they completed a problem-solving task within a GenAI-assisted intelligent tutoring system. Students' purpose of using GenAI was also analyzed from the perspective of information processing, i.e., information acquisition and information transformation. Using the sequential and clustering analysis, this study classified participants into two groups based on their SRL sequences. These two groups differed in the frequency and temporal characteristics of GenAI use. As well, most of the students used GenAI for information acquisition rather than information transformation, while the correlation between the purpose of using GenAI and learning performance was not statistically significant. Our findings informed both the pedagogical design and the development of GenAI-assisted learning environments.
\end{abstract}

\ccsdesc[500]{Applied computing~Computer-assisted instruction}

\keywords{Generative AI, self-regulated learning, intelligent tutoring system, transition}


\maketitle

\section{Introduction}
Generative artificial intelligence (GenAI) techniques, particularly those powered by large language models, have been increasingly applied in educational contexts to enhance students’ learning experiences through adaptive content generation, immediate feedback, and personalized scaffolding ~\cite{yan2024promises}. A substantial body of empirical research has demonstrated the positive effects of GenAI on learners’ self-efficacy, strategy use, and academic achievement ~\cite{ji2025does,ng2024empowering,pahi2024enhancing}. However, the integration of GenAI into traditional learning environments does not automatically ensure academic success. The ways in which students choose to interact with GenAI, such as the prompts they generate and the strategies they use to regulate these interactions, play a crucial role in shaping learning outcomes. In this regard, learners’ self-regulated learning (SRL) capacities are central to determining the quality of student–GenAI collaboration.

SRL is commonly defined as a dynamic and recursive process in which learners monitor and regulate their cognitive, affective, and behavioral processes to achieve specific learning goals ~\cite{winne2019paradigmatic,winne2019nstudy,zimmerman2000attaining}. Within GenAI-assisted learning environments, students may exhibit distinct patterns in how they regulate their use of GenAI and incorporate it into their SRL activities. For example, researchers have found that some students leverage GenAI tools to support metacognitive monitoring and regulation, two core subprocesses of SRL, by generating real-time feedback and suggestions ~\cite{wang2025optimizing}. Similarly, a study by ~\cite{kim2025students} reported that learners used GenAI to deepen their understanding of instructional materials by prompting the system to provide elaborate explanations and illustrative examples. 

Beyond these emerging findings, research remains limited in capturing the temporal dynamics of students’ SRL patterns when engaging with GenAI. Advances in learning analytics, however, offer promising opportunities to trace and characterize the dynamic nature of students’ interactions with GenAI during SRL processes. Building on this potential, the present study employed sequential clustering algorithms to identify students’ interaction patterns with GenAI in a computer-based nutrition problem-solving task. Additionally, we distinguish between two primary purposes of GenAI use: information acquisition (i.e., deepening understanding of unfamiliar information) and information transformation (i.e., evaluating or confirming information), drawing on perspectives from information processing ~\cite{anderson2014human}. The following sections outline the theoretical framework that underpins this study.

\section{Background and Related Work}

\subsection{SRL Theoretical Framework}
 Over the decades, extensive research has demonstrated that learners actively engage in their own learning processes. To reach their learning goals, they take the initiative by setting objectives and plans, choosing learning strategies, monitoring and adjusting their progress, and reflecting on their outcomes. This active engagement and goal-directed behavior constitute the core of the theoretical framework known as self-regulated learning (SRL). One prominent SRL framework is Zimmerman’s three-phase model~\cite{zimmerman2000attaining}, which consists of forethought, performance, and reflection. Each of these phases has an integral role for students in terms of how they control and improve their learning experiences. 

In the first phase of forethought, students undertake a series of preparatory actions, including setting achievable goals and plans for themselves. Specifically, they begin with analyzing the requirements of the task and identifying learning goals by clarifying expectations, understanding criteria for success, and breaking down complex assignments into manageable steps ~\cite{brady2024beyond,muljana2023examining}. They further prioritize steps, create schedules, and allocate time for each activity ~\cite{omarchevska2025flipped} as well as gather and organize necessary materials ~\cite{fischer2023using}. These cognitive processes are essential for transitioning to the next phase of performance, where they actively engage in SRL cognitive strategies for executing the task, such as task strategies of information processing (e.g., summarization or elaboration of material) or help-seeking from peers or resources when encountering difficulties ~\cite{aleven2016help,gonida2019help}. Importantly, during the performance phase, learners continuously monitor their cognitive activities and outcomes and evaluate the quality of these processes against their goals to generate in-the-moment feedback ~\cite{agustianto2016design,dinsmore2008focusing,muijs2020metacognition}. This feedback serves as a bridge to the reflection phase. During self-reflection, the feedback prompts students to review their performance and reflect on the learning process. Specifically, they attribute their outcomes and assess how effectively they acted on them~\cite{li2018effects}. This process fosters a sense of satisfaction or highlights areas for improvement, directly influencing their motivation and regulatory reactions in future learning cycles. This reflective process has both immediate and distal impacts. It enables learners to make real-time adjustments to their cognitive activities to improve efficiency and deepen their conceptual understanding ~\cite{rakovic2022examining}. Over time, this cyclical process enhances self-awareness and supports the continuous refinement of learning skills

\subsection{SRL Sequences}

Critically, SRL is not simply a static set of mental abilities; it is a dynamic, cyclical process by which learners perform a set of events or actions ~\cite{maldonado2018mining}. In acknowledging this dynamic nature of SRL, researchers has converted to a process-oriented approach to analyze SRL temporal unfolding ~\cite{saint2022temporally}. The majority of process-oriented studies relied on making inferences from trace data. Trace data refers to the high-frequency, time-stamped records of every interaction a learner has with a digital system, including clickstreams, resource access patterns, time spent on pages, video interactions, and forum posts ~\cite{gasevic2017detecting}. This methodology offers the promise of capturing a granular, objective, and unobtrusive record of learner behavior as it happens in real-time.   

Advanced analytical methods such as process mining, Markov models, Bayesian modeling, and sequence mining are used to identify and characterize the order, frequency, and transitions of actions over time ~\cite{saint2020trace}. These approaches reveal variations in how learners' behavior unfolds dynamically, distinguishing different tactics and strategies employed by learners. Notably, these temporal differences are often linked to distinct performance outcomes. For instance, ~\cite{ye2022analysis} using lag‑sequence analysis in an online learning environment, identified three learner groups based on their action sequences: a high‑performing “cognitive‑oriented” group that followed a goal‑driven pattern of active planning and deep cognitive operations; an intermediate “reflective‑oriented” group that engaged in iterative, feedback‑driven trial‑and‑error cycles; a low‑performing group with a “negative‑regulated” pattern characterized by a passive, minimally planned sequence. Similarly, an Ordered Network Analysis by ~\cite{zhao2025effect} further showed that specific transitions between SRL events (e.g., planning to monitoring) exerted a disproportionate influence on learning outcomes than others. Critically, the influence of these transitions varied across temporal stages of a task. These findings provide evidence that the sequence of strategic SRL action is a critical determinant of effective learning.

\subsection{Learning with Generative AI }
GenAI technologies, such as ChatGPT, have been widely used in the educational domain, especially to support learning across different educational levels. Since ChatGPT was released in late 2022, a large body of academic research has incorporated GenAI technologies into learning environments to foster students' learning performance and outcomes. Although previous research shared different perspectives on the impacts of GenAI use, the continued exploration of its application reflected the current development trend and attitude toward GenAI. Integrating GenAI into learning systems to provide more personalized learning experiences is a common research topic. For example, ~\cite{santhosh2024gaze} integrated ChatGPT with an adaptive learning system to enhance student engagement and learning outcomes. The results showed that students with AI-driven support achieved greater engagement, better comprehension, and improved learning outcomes. In addition, ~\cite{banjade2024empowering} developed a personalized tutoring system supported by GenAI, whereby the Ask AI Tutor function plays the role of tutors to interact with students and answer their questions during the learning process. GPTutor, developed by  ~\cite{lui2024gptutor}, were used to support students' learning processes by enabling interactive exploration of learning materials, such as summarizing learning contents, creating review exercises, and providing a conversation interface to answer questions. Additionally, Duolingo used GenAI to create interactive dialogue scenarios and provide simulated conversations for language learning ~\cite{mittal2024comprehensive}. Such studies aimed to use GenAI to provide a better learning experience and support to students, and help them achieve higher academic goals. 

As well, several studies ~\cite{afzaal2024informative,goslen2025llm,li2023effects,wang2024understanding} have been conducted to evaluate students' SRL in a GenAI-assisted learning environment and explore whether students can adapt to this new technology. A review by ~\cite{qi2025systematic} summarized how GenAI supports the distinct phases of SRL in learning activities. In the forethought phase, students employed GenAI to generate more task-related information and provide procedural guidance on how to solve the learning task. In the performance phase, students mainly use GenAI to scaffold learning activities, including summarizing, organizing, and packing information. In the phase of reflection, GenAI can be used to provide personalized feedback and potential improvement, facilitatingting the next circle of SRL. Accordingly, this study aimed to investigate students' GenAI use patterns across these three SRL phases.

\subsection{Present Study and Research Questions}
Although previous studies have integrated GenAI technologies into supporting learning, little is known about how different students use GenAI during the learning process and why they asked GenAI for help. To address these questions, this study aims to answer the following two research questions: RQ1: To what extent can students be clustered into different groups based on their SRL sequences within an a GenAI-based learning platform? If so, how do these groups differ in transitional sequences during SRL and task performance? RQ2: What is the purpose of using GenAI?

\section{Methods}

\subsection{Participants}

We recruited 114 students from a Northeast U.S. university after securing approval from the Institutional Review Board. Among these participants, 38.6\% were White/Caucasian students (N = 44), 37.7\% were Asian (N = 43), 14.9\% were Hispanic (N = 17), and 7.9\% were Black or African American (N = 9). In addition, one participant self-identified as both Asian and White and Pacific Islander. The participants' average age was 21.72 years {(\itshape SD = 3.69)}. Besides two participants who self-identified as nonbinary, 74 participants identified as female, and 38 participants identified as male. All participants signed and submitted consent forms. A sample of 110 participants was included in the investigation. In this study, 62 of them used the Ask AI function at least once.

\subsection{Learning Context and Tasks}
The Healthy Choice platform is an intelligent tutoring system (ITS) that was designed to enhance individual learners' nutrition literacy. It offers immersive experiences that support learners in developing nutrition knowledge and informed decision-making~\cite{li2025theory}. Moreover, this platform integrated the ChatGPT to provide AI assistance. The "Ask AI" function adapts to learners' unique needs, providing instant information and learning support, including nutritional knowledge and relevant resources.

There are several scenarios in this system, ranging from selecting energy drinks for studying to picking sports drinks for athletic events. In this study, participants were asked to solve nutrition recommendation tasks. Students need to read the case description to find key information and set goals for their tasks. After that, participants were presented with various real-world items (e.g., food and drinks) and asked to select the optimal choice. When participants reviewed each provided item, they could make an initial assessment on its appropriateness by checking the detailed description, such as its ingredients and calories. Students were required to submit a final decision from all possible items.

\subsection{Measurement}
To explore students' GenAI use patterns, we extracted participants' SRL activities from the trace data automatically generated by the Healthy Choice platform, which recorded students' interactive activities in this platform. Guided by Zimmerman’s cyclical phase model~\cite{zimmerman2000attaining}, we developed the following coding scheme including both the macro-level SRL phases and micro-level SRL activities (see Table~\ref{tab:code}).


\vspace{1em}
\begin{table}[ht]
\centering
\caption{The SRL Activities from Healthy Choice Logfiles}
\label{tab:code}

\begin{tabularx}{\linewidth}{
  >{\raggedright\arraybackslash}p{0.18\linewidth}  
  >{\raggedright\arraybackslash}p{0.25\linewidth}  
  >{\raggedright\arraybackslash}p{0.16\linewidth}  
  >{\raggedright\arraybackslash}p{0.35\linewidth}  
}
\toprule
\textbf{Macro-level SRL} & \textbf{Micro-level SRL (Actions)} & \textbf{Details} & \textbf{Description} \\
\midrule

\multirow[t]{2}{*}{Forethought}
  & \multirow[t]{2}{*}{Goal-Setting (GS)} & Add keywords     &Students set task-specific (sub)goals from the scenario descriptions by highlighting the important information or keywords.  \\
  &                                       & Delete keywords  &  \\

\multirow[t]{7}{*}{Performance}
  & Execution (EX)        & Check an item       &Students checked the detailed information of the ten provided items one by one.  \\
  & Evaluation (EV)       & Choose a label      &Students evaluated whether the specific item was appropriate for solving the problem, considering the specific scenario.  \\
  & Orientation (OR)      & Initial decision    &Students made an initial decision on whether to select a specific item for further consideration or not.  \\
  & \multirow[t]{2}{*}{Monitoring (MO)} & Rank items       &  \\
  &                         & Delete items       &  \\
  & Elaboration (EL)      & Compare items       &Students synthesized and compared different pieces of information of selected items to form a final decision.  \\
  & Submission (SU)       & Submit decisions    &Students submitted the final item.  \\

\multirow[t]{2}{*}{Self-reflection}
  & \multirow[t]{2}{*}{Reflection (RE)} & Save summary     &Students reflected on the task trajectory by summarizing how they achieved the final decision.  \\
  &                                      & Submit summary   &  \\
\bottomrule
\end{tabularx}
\end{table}

For the prompts in the Ask AI, we classified them based on the perspective of information processing into two types: information acquisition and information transformation, as shown in Table~\ref{tab:type}. Two coders coded the prompts and compared the results, achieving  an agreement of 74\%. The inconsistent coded data were discussed again and resolved through a voting process conducted by the two coders and a third coder.

\vspace{1em}
\begin{table}[ht]
\centering
\caption{Classifying the Prompts Based on the Perspective of Information Processing}
\label{tab:type}

\begin{tabularx}{\linewidth}{
  >{\raggedright\arraybackslash}p{0.22\linewidth}
  >{\raggedright\arraybackslash}p{0.44\linewidth}
  >{\raggedright\arraybackslash}p{0.28\linewidth}
}
\toprule
\textbf{Type} & \textbf{Descriptions} & \textbf{Examples} \\
\midrule

Information acquisition
&
Students interacted with AI for the purpose of a general understanding of unfamiliar information, such as specific terms, effects, etc.
&
``How much mg caffeine will cause anxiety in an adult male'' \\
& & ``Effects of sodium and potassium'' \\
\midrule

Information transformation
&
Students requested assistance from AI to help them evaluate or confirm some prior knowledge or resolve cognitive conflicts.
&
``If an energy drink contains carbohydrate, is it still sugar-free?'' \\
& & ``What’s the main difference between body armor drink and body armor super hydration?'' \\

\bottomrule
\end{tabularx}
\end{table}

\subsection{Data Analysis}
We conducted the data analysis using R through R studio and Python through Google Colab. We employed sequential and clustering analysis to classify participants into different groups based on their SRL transition sequences. The analysis includes two-step process. First, we created a state sequence from the action variables. Second, we built the clusters based on the identified sequences using hierarchical clustering algorithms
~\cite{huang2025modeling}. This aims to identify participants' learning behavior during the tasks and find their purposes on using Ask AI. In addition, we conducted transition rate heatmaps to compare clusters and examine their performance differences.

\begin{figure}[t!]
    \centering
    \includegraphics[width=0.85\columnwidth]{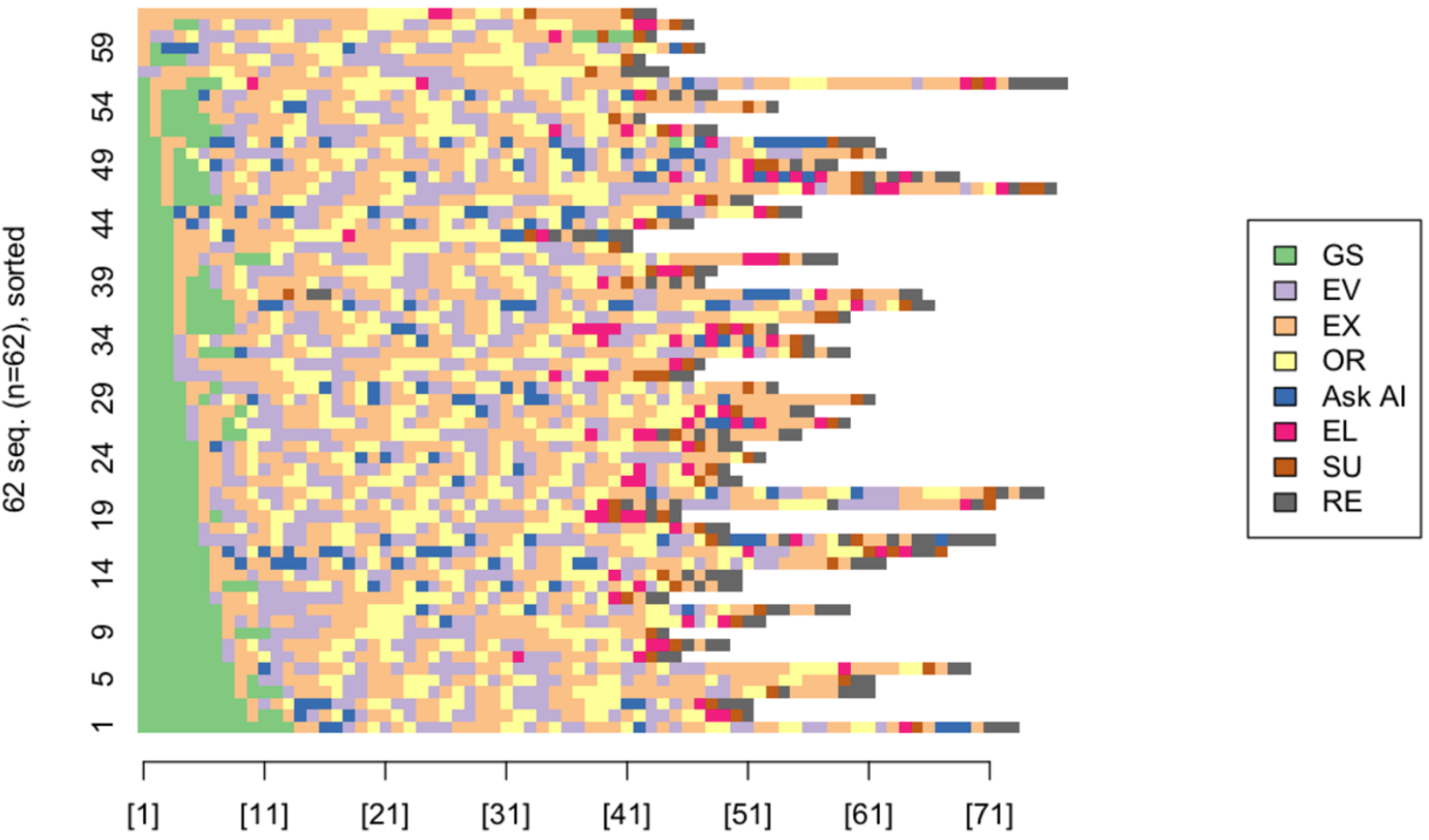}
    \caption{All Participants' SRL Transition Sequences.}
    \vspace{-10pt}
    \label{Result}
\end{figure}

\begin{figure}[!htbp]
  \centering
  \includegraphics[scale=0.20]{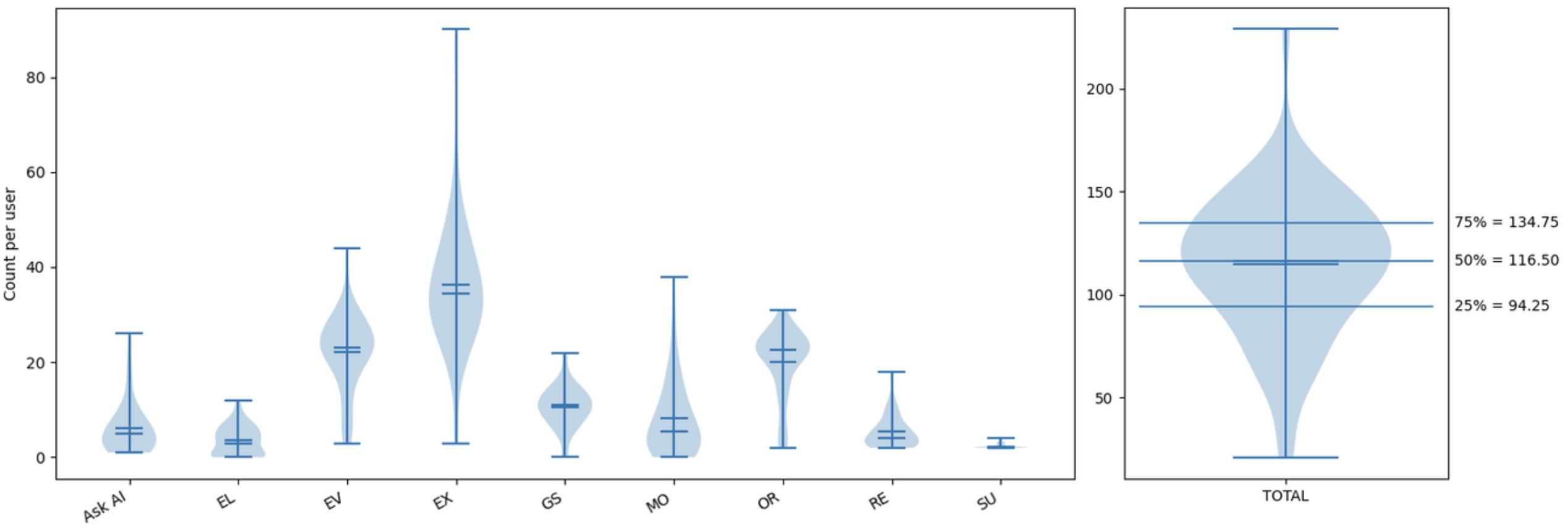}
  \caption{Distribution of Action Count per Participant.}
  \Description{action count.}
  \label{action}
\end{figure}

\section{Results}
\subsection{SRL Sequences}

Figure\ref{Result} shows the transitional sequences of 62 participants while taking the Ask AI activity at least once during the tasks. Most of the participants (N = 55, 88.7\%) started the task from GS and others started from EX (N = 7, 11.3\%). The number of actions taken by participants during the task ranged from 21 to 229 (SD = 34.8). Figure\ref{action} illustrates the distribution of action counts among participants. A majority of participants (N = 40) took Ask AI activities between 1 and 6 times (M = 6), followed by 7-20 times (N = 20) and more than 20 times (N = 2). 

\begin{figure}[!htbp]
  \centering
  \includegraphics[scale=0.28]{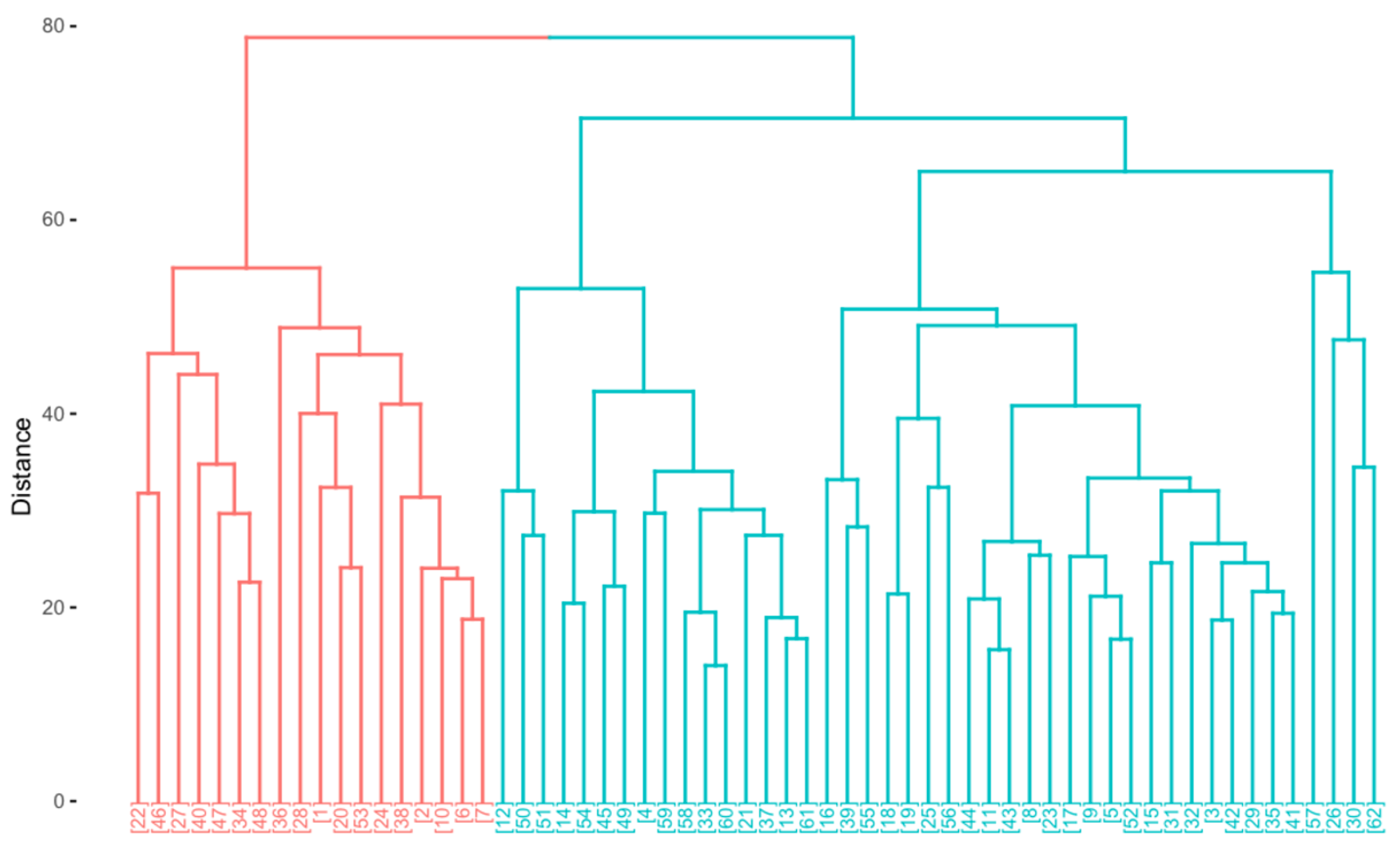}
  \caption{Dendrogram of SRL Sequences.}
  \Description{Dendrogram of SRL Sequences.}
  \label{cl1}
\end{figure}

\begin{figure}[!htbp]
  \centering
  \includegraphics[scale=0.20]{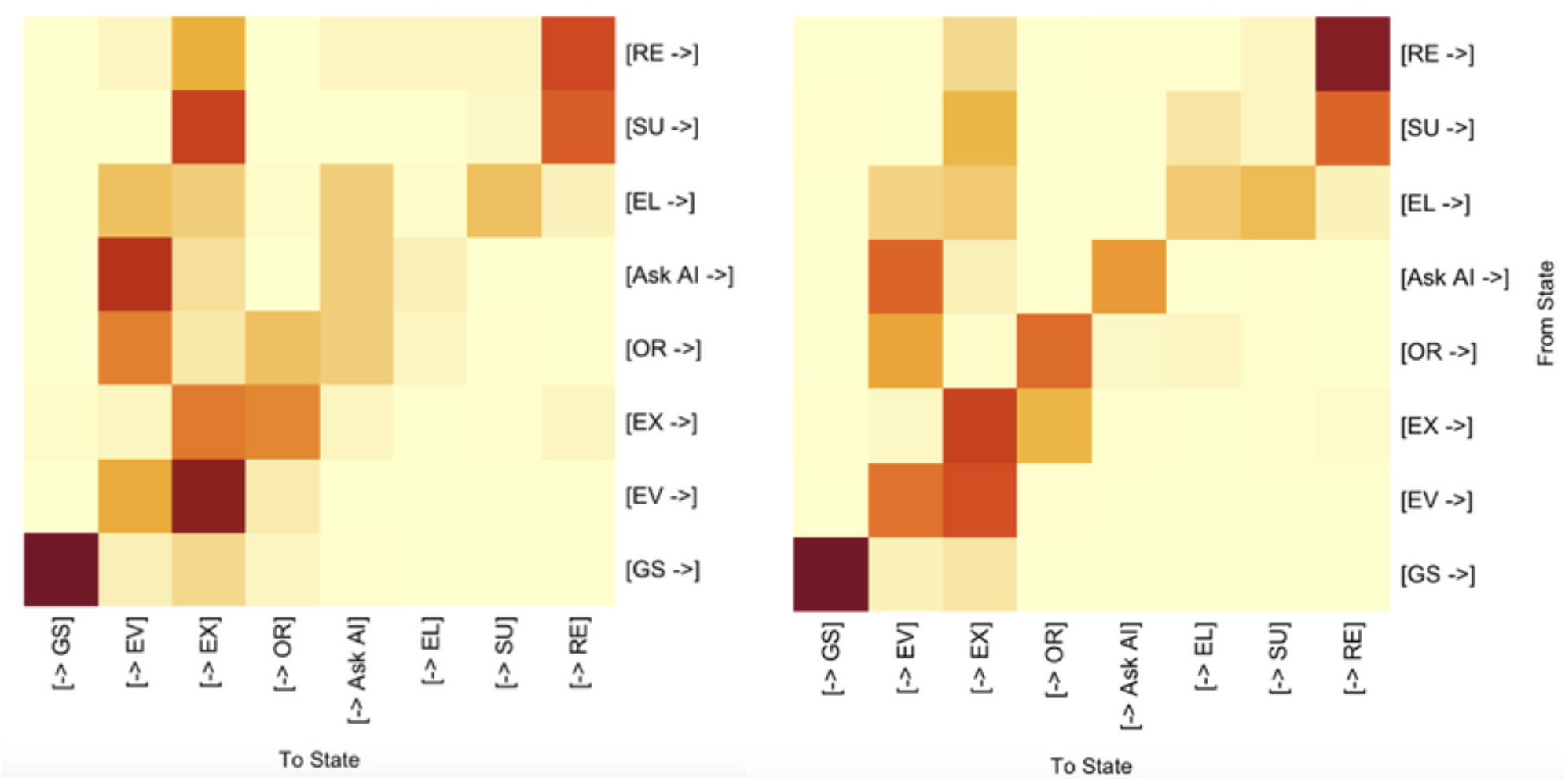}
  \caption{Transition Rate Heatmap. Cluster 1 (left) and Cluster 2 (right).}
  \Description{Transition Rate Heatmap.}
  \label{cl3}
\end{figure}

We used the Elbow method to determine the optimal number of clusters and then classified participants into two groups. According to Figure\ref{cl1}, we classified the participants into two groups based on their SRL sequences within the Healthy Choice platform. Cluster 1 includes 18 participants and cluster 2 includes 44 participants. Furthermore, we analyzed and compared the clusters' transition rate through heatmaps (see Figure\ref{cl3}). The heatmaps revealed different sequences patterns among participants and groups. In cluster 1,the most prominent transition was EV$\rightarrow$EX, indicating a process of moving from goal setting to monitoring progress and performing tasks. In addition, significant transitions were observed for Ask AI$\rightarrow$EV, SU$\rightarrow$EX, and SU$\rightarrow$RE, followed by OR$\rightarrow$EV and EX$\rightarrow$OR. In cluster 2, significant transitions were observed for EV$\rightarrow$EX, SU$\rightarrow$RE, and Ask AI$\rightarrow$EV. Compared to cluster 1, SU$\rightarrow$EX had a lower transition rate. Notably, the transition rate on EX$\rightarrow$OR was slightly lower than that of cluster 1. Although both clusters had similar patterns, cluster 1 had higher transition rates on several sequences, such as SU$\rightarrow$EX, Ask AI$\rightarrow$EV, EV$\rightarrow$EX.

\begin{figure}[!htbp]
  \centering
  \includegraphics[scale=0.28]{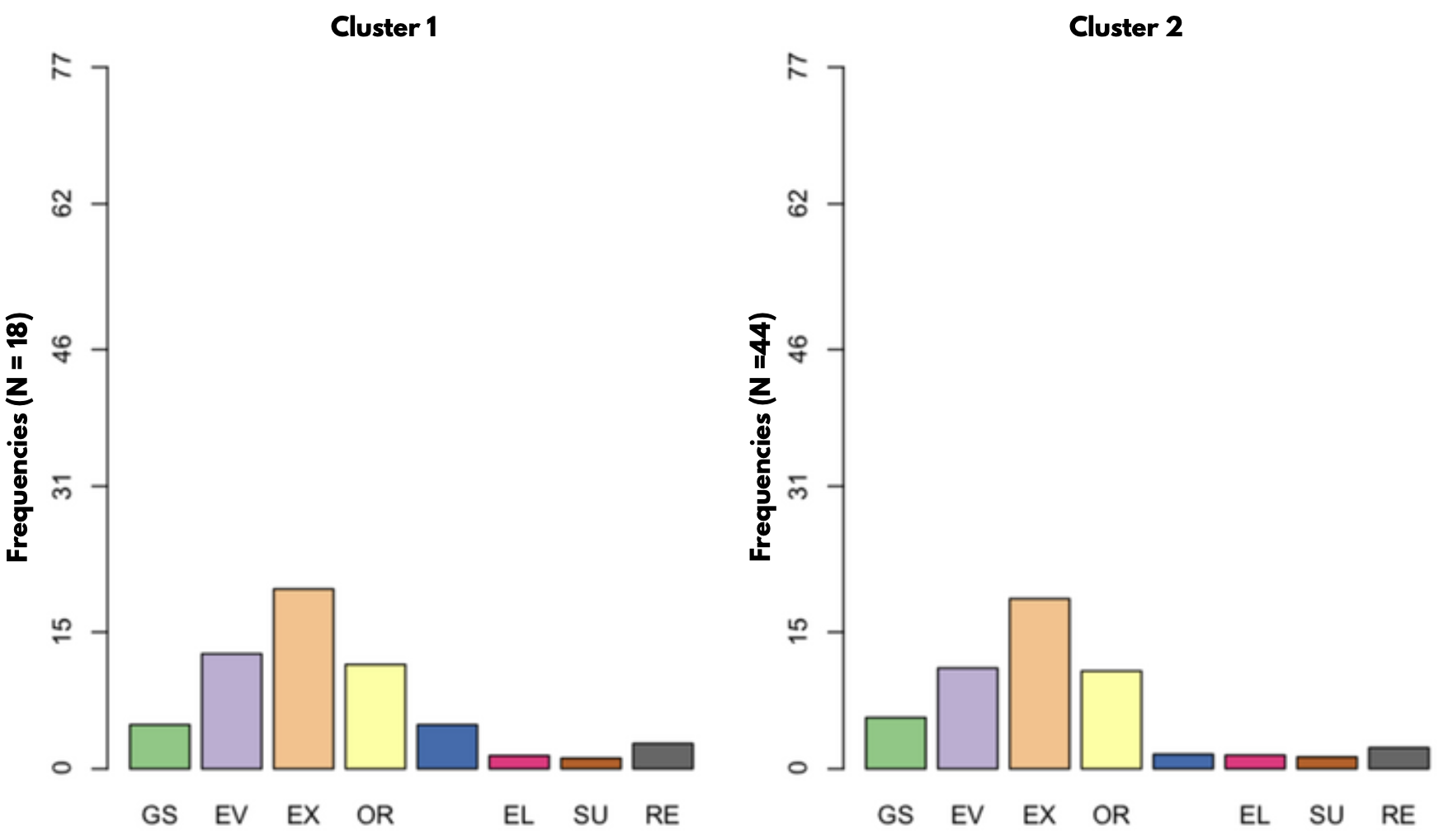}
  \caption{Action Frequency of Each Cluster (Blue represents Ask AI).}
  \Description{Action Frequence of Each Cluster .}
  \label{cl2}
\end{figure}

Figure\ref{cl2} shows the action frequencies of each cluster. Both clusters have similar distributions, except for the frequency of Ask AI. Obviously, students in cluster 1 used GenAI for help more often (5 times) than those in cluster 2 (2 times).

\subsection{Purpose of Using GenAI}
A total of 385 prompts were classified into two types: information acquisition (IA) and information transformation (IT). According to the analysis, students used GenAI more often for IA (N = 204) than IT (N = 172). The Wilson confidence intervals were [49.20\%, 59.22\%] for IA and [40.78\%, 50.80\%] for IT. The observed difference (8.52\%) suggests that students used GenAI with the IA purpose, such as information retrieval and requesting additional details, at a higher prevalence.

\section{Discussion and Conclusions}
  
This study applied a sequential clustering algorithm to uncover students’ dynamic SRL patterns within a GenAI-assisted learning environment. The analysis revealed two distinct clusters, differentiated by the transition sequences of students’ self-regulated problem-solving activities, including their interactions with GenAI. With respect to the transition sequences, this study identified three key differences between Cluster 1 and Cluster 2. As illustrated by the diagonals in Figure\ref{cl3}, students in Cluster 1 were more likely to engage in consecutive goal-setting (GS$\rightarrow$GS) and execution (EX$\rightarrow$EX) activities, whereas students in Cluster 2 tended to repeat extended sequences of goal-setting (GS$\rightarrow$GS) and self-reflection (RE$\rightarrow$RE) activities. Notably, self-reflection is recognized as a critical learning strategy for expertise development, as it facilitates the construction and updates of cognitive schemas relevant to the learning task. 

Furthermore, students in Cluster 1 frequently transitioned from evaluation to execution (EV$\rightarrow$EX), reflecting active engagement in gathering and assessing evidence to solve the task. However, this pattern also suggests that these students spent more time analyzing individual pieces of information rather than organizing the collected information to construct a coherent mental representation of the problem-solving task, which may have contributed to their lower performance. Additionally, the analysis of students’ integration of GenAI tools into SRL processes revealed distinct patterns. Cluster 1 transitioned from requesting help from GenAI to evaluation (Ask AI$\rightarrow$EV), indicating that they primarily used AI to clarify unfamiliar information before evaluating it. However, Cluster 2 exhibited repeated sequences involving Ask AI, reflecting deeper interactions with GenAI through multiple rounds of dialogue.

Additionally, this study distinguished between two GenAI usage purposes based on information processing: information acquisition and information transformation. While students used GenAI more for acquisition (e.g., understanding new terms) than for transformation (e.g., organizing or evaluating information), the latter is a higher-order cognitive process more critical for developing metacognitive skills and achieving learning outcomes. The infrequency of such transformative use highlights an imperative for educators: to design interventions that scaffold students' ability to use GenAI for more than just information lookup, thereby fostering its deeper educational benefits.

The findings of this study have significant implications for both pedagogy and the design of technology-enhanced learning environments. First, the results highlight the critical role of self-reflection in supporting effective problem-solving, suggesting that instructional designs should explicitly foster iterative reflection cycles through scaffolds or prompts that guide students in evaluating and integrating their goals, actions, and outcomes. Second, the observed patterns of GenAI use indicate that both the depth and sequencing of AI interactions influence learning outcomes. Students who engaged in multiple rounds of dialogue with GenAI exhibited more accurate problem-solving, underscoring the potential of adaptive AI tools that support sustained and structured interactions aligned with SRL strategies. Third, the findings highlight the need for targeted interventions for students exhibiting less effective SRL patterns, such as those who focus on isolated information analysis without integrating it into a coherent mental representation. Future research could examine how real-time learning analytics can be incorporated into GenAI-supported environments to provide personalized guidance and promote more effective problem-solving strategies.



\bibliographystyle{ACM-Reference-Format}
\bibliography{Reference}
\end{document}